# Receipt of hybrid entangled and CV entangled states on demand

Sergey A. Podoshvedov[1] and Mikhail S. Podoshvedov[2]

[1]*Laboratory of Quantum Information Processing and Quantum Computing, Institute of Natural and Exact Sciences, South Ural State University (SUSU), Lenin Av. 76, Chelyabinsk, Russia*
[2]*Institute of Physics, Kazan Federal University (KFU), 16a Kremlyovskaya St., Kazan, Russia*
email: *sapodo68@gmail.com*[1], *mikepodo6@gmail.com*[2]

**Abstract.** We propose a new approach to the generation of entangled states, both hybrid and consisting exclusively of continuous variable (CV) states. A single mode squeezed vacuum is mixed with a delocalized single photon on arbitrary beam splitter (BS) with subsequent registration of some measurement outcome in auxiliary mode. The entangled states are generated whenever any event is measured in auxiliary mode. Negativity is used as a measure of entanglement. Under certain initial set conditions, the conditioned state becomes as entangled as possible. New types of CV states are introduced. This approach can be expanded to implement a high complexity quantum network.

Entanglement is a property of different physical systems, even being separated by distance, to be exclusively described by one wave function (state). The strange property forms the basis for the most remarkable, purely quantum effects and is the key ingredient for many applications of quantum mechanics in both discrete variable (DV) or the same qubit and continuous variables (CV) regimes. Although, initially the perfect correlations (Einstein-Podolsky-Rosen (EPR) were used by trio of collaborators [1] to define "element of reality" a notion, according to them, which must certainly be present in all quantum theories. The concept of the "element of reality" was staggered when Bell derived his remarkable inequalities [2] to test quantum description of two-partite entangled system. The pioneering test of Bell's inequality was done by Freedman and Clauser [3] followed by the famous "Bell experiment" works [4-7]. Loophole-free Bell inequality violation was demonstrated in [8]. The measurement settings were determined from photons coming from distant quasars in order to exclude "freedom-of-choice" loophole in testing quantum mechanics [9]. In the cases, local realism and quantum mechanics show completely opposite predictions.

The standard source of entangled-photon pairs is the nonlinear optical process of spontaneous parametric down-conversion (SPDC) [10]. Now, SPDC sources of entangled photon pairs of high quality can be routinely realized using various methods [11-13]. Nearest generalization of the EPR state is Greenberger-Horne Zeilinger (GHZ) state [14] that involves at least three subsystems and shows more powerful demonstration of the quantum nonlocality [15]. Cluster states [16] attract much attention for one-way quantum computation due to the multi-particle entanglement [17,18]. Another type of entanglement can be detected in entangled states formed by coherent states with displacement amplitudes opposite in sign [19]. Introduction of the entangled state is based on Schrödinger cat state thought paradox [20] which was initially called up to illustrate weakness of the Copenhagen interpretation of quantum mechanics applied to everyday objects. Now optical version of the Schrödinger cat state (SCS) well, at least theoretically, studied, although it is an incredibly complex implementation object especially if coherent states take on values $\geq 2$ [21]. A natural generalization of the entangled coherent states is the hybrid entangled states (namely, in such terms the gedanken cat was initially formulated [20]) in which entanglement is formed by objects of various physical nature [22-26]. The potential of such states for quantum information processing is quite high [27-30].

However, at that time, no effective sources of entanglement with three-, four and more photons are present, although the entanglement with more than two particles can be full of new exiting classically paradoxical phenomena. Thus far, the task of implementing the entangled states is a challenge for researchers. In the vast majority of cases, the problem of receipt of the entangled states different from EPR one is reduced to the building of complex optical schemes, the use of significant resources, and that the most important, the conditional generation due to the observation of rare measurement outcome in auxiliary modes. In addition, at the initial stage, it is advisable not to use exotic states that are hard to implement in advance. Our goal is to remove the obstacles and propose a new approach to obtain entangled both hybrid (in our case, physical objects from different Hilbert spaces) and CV states on demand.

The input elements for the optical scheme in Fig. 1 are a single mode squeezed vacuum (SMSV) [31] and a single photon. SMSV is routinely produced in laboratories and can be represented as

$$|SMSV\rangle = S(r)|0\rangle = \sum_{n=0}^{\infty} b_{2n}|2n\rangle = \frac{1}{\sqrt{\cosh r}} \sum_{n=0}^{\infty} (\tanh r)^n \frac{\sqrt{(2n)!}}{2^n n!} |2n\rangle, \quad (1)$$

where of squeeze operator $S(r) = \exp(r(a^2 - a^{+2})/2)$ is introduced, where for simplicity we assumed the squeezing parameter $r$ to be real with $a$ and $a^+$ being annihilation and creation operators. The single photon can be conditionally created by registering a single photon in one of two modes of SPDC.

To generate the optical entangled hybridity (macro-micro entanglement), we are also going to use a delocalized photon

$$|\varphi\rangle_{23} = a_0|01\rangle_{23} + a_1|10\rangle_{23}, \quad (2)$$

occupying simultaneously modes 2 and 3, where the amplitudes $a_0$ and $a_1$ satisfy the normalization condition $|a_0|^2 + |a_1|^2 = 1$. A single photon can be easily converted to delocalized by passing it through the beam splitter. In essence, the amplitudes $a_0$ and $a_1$ are simply the parameters of the BS, namely, the transmittance and reflection coefficient.

The input SMSV (1) is mixed with the delocalized photon in Eq. (2) on the BS which is described by the following unitary matrix

$$BS = \begin{bmatrix} t & -r \\ r & t \end{bmatrix}, \quad (3)$$

where $t$ and $r$ are the real transmittance and reflectance coefficients, satisfying the normalization condition $t^2 + r^2 = 1$ as shown in Fig. 1. The SMSV occupying the first mode is mixed with the second one of the delocalized photon on beam splitter mode as

$$BS_{12}(|SMSV\rangle_1|\varphi\rangle_{23}) = a_0 BS_{12}(|SMSV\rangle_1|0\rangle_2)|1\rangle_3 + a_1 BS_{12}(|SMSV\rangle_1|1\rangle_2)|0\rangle_3, \quad (4)$$

due to linearity of the BS unitary operator. The conditional state is generated in the case of registration of either an even number $p = 2m$ or an odd number $p = 2m + 1$ of photons in the second auxiliary mode. In order to obtain the analytical form of the conditional state, one needs to follow the unitary evolution of the following terms $BS_{12}(|SMSV\rangle_1|0\rangle_2)$ and $BS_{12}(|SMSV\rangle_1|1\rangle_2)$, separately. Since the creation operators are transformed as $a_1^+ \to ta_1^+ - ra_2^+$ and $a_2^+ \to ra_1^+ + ta_2^+$, we can finally obtain the output states being of result of mixing arbitrary $l$ Fock state $|l\rangle_1$ with vacuum $|0\rangle_2$

$$BS_{12}(|l\rangle_1|0\rangle_2) = \sum_{k=0}^{l} (-1)^k t^{l-k} r^k \sqrt{\frac{l!}{k!(l-k)!}} |l-k\rangle_1|k\rangle_2, \quad (5)$$

and arbitrary $l$ Fock state $|l\rangle_1$ with the single photon $|1\rangle_2$

$$BS_{12}(|l\rangle_1|1\rangle_2) = \sqrt{l+1} t^l r |l+1\rangle_1|0\rangle_2 +$$
$$\sum_{k=0}^{l} (-1)^k \frac{t^{l-k-1}r^k}{k!} \sqrt{\frac{(k+1)!l!}{(l-k)!}} \left(t^2 - \frac{l-k}{k+1}r^2\right) |l-k\rangle_1|k\rangle_2. \quad (6)$$

Considering the fact that the SMSV contains only even number states and collecting all terms with either an even $|p = 2m\rangle_2$ or odd $|p = 2m+1\rangle_2$ number of photons in the second auxiliary mode, it can be shown that the following conditional entangled hybrid state is generated

$$|\Delta_0\rangle_{13} = N_0(a_0|\Psi_0\rangle_1|1\rangle_3 + a_1 B_0|\Phi_0\rangle_1|0\rangle_3), \quad (7)$$

where the CV states are given by

$$|\Psi_0\rangle = L_0 \sum_{k=0}^{\infty} b_{2k} t^{2k} |2k\rangle, \tag{8}$$

$$|\Phi_0\rangle = K_0 \sum_{k=0}^{\infty} b_{2k} t^{2k} \sqrt{2k+1} |2k+1\rangle, \tag{9}$$

provided that vacuum state $|0\rangle_2$ is fixed in second auxiliary mode, where the amplitudes $b_{2k}$ are the SMSV ones given in Eq. (1), while the $L_0 = (\sum_{k=0}^{\infty} |b_{2k}|^2 |t|^{4k})^{-1/2}$ and $K_0 = (\sum_{k=0}^{\infty} |b_{2k}|^2 |t|^{4k} (2k+1))^{-1/2}$ are the normalization factors of the CV states $|\Psi_0\rangle$ and $|\Phi_0\rangle$, respectively, and $N_0 = (|a_0|^2 + |a_1|^2 |B_0|^2)^{-1/2}$ is the normalization factor of the hybrid entangled state $|\Delta_0\rangle_{13}$ in Eq. (7). The parameter $B_0$ follows from direct calculation

$$B_0 = \frac{rL_0}{K_0}. \tag{10}$$

The success probability to conditionally generate the entangled hybrid state (7) is given by

$$P_0 = \frac{1}{L_0^2 N_0^2}. \tag{11}$$

Note that the state $|\Psi_0\rangle$ contains exclusively even Fock states, while the state $|\Phi_0\rangle$ is a superposition of odd Fock states. Therefore, they can be called even and odd CV states, respectively.

Suppose that an even number of photons $p = 2m$ is registered in the second auxiliary mode. Then, the following hybrid entangled state is created

$$|\Delta_{2m}\rangle_{13} = N_{2m}(a_0 |\Psi_{2m}\rangle_1 |1\rangle_3 - a_1 B_{2m} |\Phi_{2m}\rangle_1 |0\rangle_3), \tag{12}$$

where even CV state $|\Psi_{2m}\rangle$ is given by

$$|\Psi_{2m}\rangle = L_{2m} \sum_{k=0}^{\infty} b_{2(k+m)} t^{2k} \sqrt{\frac{(2(k+m))!}{(2k)!}} |2k\rangle, \tag{13}$$

while odd CV state is written as

$$|\Phi_{2m}\rangle = K_{2m} \sum_{k=0}^{\infty} b_{2(k+m)} t^{2k} \sqrt{\frac{(2(k+m))!}{(2k+1)!}} \left(t^2 - \frac{2k+1}{2m} r^2\right) |2k+1\rangle, \tag{14}$$

with the normalization factors $L_{2m} = \left(\sum_{k=0}^{\infty} |b_{2(k+m)}|^2 |t|^{4k} \frac{(2(k+m))!}{(2k)!}\right)^{-1/2}$ and $K_{2m} = \left(\sum_{k=0}^{\infty} |b_{2(k+m)}|^2 |t|^{4k} \frac{(2(k+m))!}{(2k+1)!} \left|t^2 - \frac{2k+1}{2m} r^2\right|^2\right)^{-1/2}$, respectively. Normalization factor of the hybrid entangled state $|\Delta_{2m}\rangle_{13}$ is the following $N_{2m} = (|a_0|^2 + |a_1|^2 |B_{2m}|^2)^{-1/2}$. Here, the additional factor $B_{2m}$ becomes

$$B_{2m} = \frac{2mL_{2m}}{rK_{2m}}. \tag{15}$$

where $m \neq 0$. The success probability to fix outcome of $2m$ photons and conditionally generate the hybrid entangled state $|\Delta_{2m}\rangle_{13}$ is defined by

$$P_{2m} = \frac{|r|^{4m}}{(2m)! L_{2m}^2 N_{2m}^2}. \tag{16}$$

If an odd number of photons $p = 2m + 1$ is detected in the second auxiliary mode, then the generated hybrid entangled state will have the form

$$|\Delta_{2m+1}\rangle_{13} = N_{2m+1}(a_0 |\Psi_{2m+1}\rangle_1 |1\rangle_3 - a_1 B_{2m+1} |\Phi_{2m+1}\rangle_1 |0\rangle_3), \tag{17}$$

with the following CV states odd $|\Psi_{2m+1}\rangle$

$$|\Psi_{2m+1}\rangle = L_{2m+1} \sum_{k=0}^{\infty} b_{2(k+m+1)} t^{2k} \sqrt{\frac{(2(k+m+1))!}{(2k+1)!}} |2k+1\rangle, \tag{18}$$

and even $|\Phi_{2m+1}\rangle$

$$|\Phi_{2m+1}\rangle = K_{2m+1} \sum_{k=0}^{\infty} b_{2(k+m)} t^{2(k-1)} \sqrt{\frac{(2(k+m))!}{(2k)!}} \left(t^2 - \frac{2k}{2m+1} r^2\right) |2k\rangle, \tag{19}$$

with the normalization factors $L_{2m+1} = \left(\sum_{k=0}^{\infty} |b_{2(k+m+1)}|^2 |t|^{4k} \frac{(2(k+m+1))!}{(2k+1)!}\right)^{-1/2}$ and $K_{2m+1} = \left(\sum_{k=0}^{\infty} |b_{2(k+m+1)}|^2 |t|^{4(k-1)} \frac{(2(k+m))!}{(2k)!} \left|t^2 - \frac{2k}{2m+1} r^2\right|^2\right)^{-1/2}$, respectively. General normalization factor of the state $|\Delta_{2m+1}\rangle_{13}$ is $N_{2m+1} = (|a_0|^2 + |a_1|^2 |B_{2m+1}|^2)^{-1/2}$ with additional factor

$$B_{2m+1} = \frac{(2m+1)L_{2m+1}}{rK_{2m+1}}. \tag{20}$$

The success probability to conditionally produce the state $|\Delta_{2m+1}\rangle_{13}$ is the following

$$P_{2m+1} = \frac{|r|^{2(2m+1)}|t|^2}{(2m+1)!L_{2m+1}^2 N_{2m+1}^2}. \tag{21}$$

Tedious calculations make it possible to verify that the success probabilities are normalized $\sum_{m=0}^{\infty}(P_{2m} + P_{2m+1}) = 1$.

As can be seen from the above expressions, the parity of the CV states either $|\Psi\rangle$ or $|\Phi\rangle$ is completely determined by the measurement outcomes in the second auxiliary mode. Nevertheless, regardless of the parity of the measurement outcomes, the CV states will always be orthogonal to each other $\langle\Phi|\Psi\rangle = 0$ since they are exclusively formed from either even or odd Fock states. The parity of the CV states depending on the parity of the Fock states measured in the second auxiliary mode is presented in Table 1. Note that the generated hybrid entangled states can be transformed $|\Delta_{2m}\rangle \to N_{2m}(a_0|\Psi_{2m}\rangle|1\rangle + a_1 B_{2m}|\Phi_{2m}\rangle|0\rangle)$, $|\Delta_{2m+1}\rangle \to N_{2m+1}(a_0|\Psi_{2m+1}\rangle|1\rangle + a_1 B_{2m+1}|\Phi_{2m+1}\rangle|0\rangle)$, respectively, by applying the phase shifter to the single photon $P|1\rangle \to -|1\rangle$.

| measurement outcome | even | odd |
|---|---|---|
| $|\Psi\rangle$ | even | odd |
| $|\Phi\rangle$ | odd | even |

**Table 1.** Generalization of the formulas (7)-(9), (12)-(14) and (17)-(19) explaining the parity of the CV states $|\Psi\rangle$ and $|\Phi\rangle$ being part of the hybrid entangled states $|\Delta_{2m}\rangle$ and $|\Delta_{2m+1}\rangle$, respectively. The parity of the CV states depends on the parity of the measurement outcome either $2m$ or $2m+1$ in auxiliary mode.

The hybrid entangled states also contain a parameter $B_{2m}$ or $B_{2m+1}$, respectively, that largely determines the measure of entanglement of the conditioned states $|\Delta_{2m}\rangle$ and $|\Delta_{2m+1}\rangle$. Indeed, the hybrid entangled states exist in the four-dimensional Hilbert space. Regardless of the squeezing parameter $r$ of input SMSV, the BS parameters and measurement outcomes, only a set of orthogonal states $\{|even\rangle_1|0\rangle_2, |odd\rangle_1|0\rangle_2, |even\rangle_1|1\rangle_2, |odd\rangle_1|1\rangle_2\}$ can be basis states of the four-dimensional Hilbert space, where by designations $|even\rangle, |odd\rangle$ it is meant the states that exclusively contain either even or odd Fock states. The state $|even\rangle$ can be either $|even\rangle = |\Psi\rangle$ or $|even\rangle = |\Phi\rangle$ as it follows from Table 1. Also, the state $|odd\rangle$ can be either $|odd\rangle = |\Psi\rangle$ or $|even\rangle = |\Phi\rangle$ with appropriate measurement outcomes (Table 1). Measure of the entangled state living in four-dimensional Hilbert space can be estimated by using partial transpose (PPT) criterion for separability [32,33] and the negativity $\mathcal{N}$ of the state can be calculated to be

$$\mathcal{N}_{2m,2m+1} = \frac{2|a_0||a_1||B_{2m,2m+1}|}{|a_0|^2 + |a_1|^2|B_{2m,2m+1}|^2}, \tag{22}$$

where the negativity value ranges from $\mathcal{N}_{min} = 0$ (separable state) up to $\mathcal{N}_{max} = 1$ (maximally entangled state). As can be seen from the expressions, negativity can take on zero value only if either $B_{2m,2m+1} = 0$ or $a_0(a_1) = 0$. The condition $a_0(a_1) = 0$ is initially impossible since this violates the delocalization condition for the initial photon in Eq. (2). The condition $B_{2m,2m+1} = 0$ is possible only in strictly limiting cases either $t = 0(r = 1)$ or $t = 1(r = 0)$ that is beyond of our consideration. Thus, the conditional state in Fig. 1 always contains some degree of the entanglement regardless of the measurement outcome in the auxiliary mode which allows us to claim that receipt of the hybrid entangled state occurs on demand. The maximum degree of the negativity is observed under the condition $|a_0| = |a_1||B_{2m,2m+1}|$. If the delocalized photon is balanced $(|a_0| = |a_1| = 1/\sqrt{2})$, then the condition of maximal negativity becomes $|B_{2m,2m+1}| = 1$.

In Fig. 2 we show the dependences of the negativity $\mathcal{N}_0$ on the squeezing parameter $r$ and transparency coefficient $t$ of the BS for various cases of $|a_1|$ of the initial delocalized photon. The set of plots is presented in order to visually assess the range of input parameters at which negativity takes on a maximum value (in the limit $\mathcal{N}_0 = \mathcal{N}_{max} = 1$). In Figure 3, we show the dependence of the success probability of generating the conditional state $|\Delta_0\rangle$ in Eq. (7) in the dependence on $r$ and $t$ at the same values of $|a_1|$ of the delocalized photon. As can be seen from the plots, there are areas $(r,t)$ in which the success probability $P_0$ takes values close to unit value $P_0 \approx 1$. It is worth noting that the success probabilities $P_1$ of receipt of the hybrid entangled state $|\Delta_1\rangle$ will already take values significantly less of one. Some values of the experimental parameters $(r, t, |a_1|)$ which provide the generation of the entangled hybrid state with the greatest possible entanglement $\mathcal{N}_0 = \mathcal{N}_{max} = 1$ are presented in Table 2. As can be seen from the Table 2, the proposed approach allows the generation of maximally entangled states with an almost deterministic success probability $P_0 \approx 1$.

| $n$ | $r$ | $t$ | $|a_1|$ | $P_0$ |
|---|---|---|---|---|
| 1 | 0.107632 | 0.423201 | 0.741004 | 0.896792 |
| 2 | 0.380541 | 0.326343 | 0.726463 | 0.88067 |
| 3 | 0.541383 | 0.259528 | 0.719131 | 0.840088 |
| 4 | 0.753348 | 0.234748 | 0.716839 | 0.7449845 |
| 5 | 0.83396 | 0.0762081 | 0.708133 | 0.728679 |
| 6 | 0.0265654 | 0.0220391 | $1/\sqrt{2}$ | 0.999404 |
| 7 | 0.303502 | 0.025593 | $1/\sqrt{2}$ | 0.955334 |
| 8 | 0.613125 | 0.020327 | $1/\sqrt{2}$ | 0.837402 |

**Table 2.** Partial experimental conditions $(r, t, |a_1|)$ under which the negativity $\mathcal{N}_0$ takes on maximal value $\mathcal{N}_0 = \mathcal{N}_{max} = 1$. The values of the success probability $P_0$ to conditionally generate the hybrid maximally entangled state are also shown. The parameters are given for both unbalanced (top five lines) and balanced $|a_1| = 1/\sqrt{2}$ (bottom three lines) delocalized photon.

Suppose the hybrid entangled state $|\Delta_0\rangle_{13}$ in Eq. (7) is already available. Let us reuse the optical scheme in Figure 1 in order to mix SMSV in Eq. (1) now occupying the second mode $|SMSV\rangle_2$ with the third DV mode of the hybrid state $|\Delta_0\rangle_{13}$ $(BS_{23}(|SMSV\rangle_2|\Delta_0\rangle_{13}))$ followed by registration of the measurement outcomes in the third auxiliary mode $M_3^{(0)}BS_{23}(|SMSV\rangle_2|\Delta_0\rangle_{13})$, where $M_3^{(0)} = |0\rangle_3\langle 0|$ is the projection operator on vacuum state. Then, using expressions in Eqs. (5)-(6) and following the same calculation technique, one can show that the next entangled state

$$|\Omega_0\rangle_{12} = a_0|\Psi_0\rangle_1|\Phi_0\rangle_2 + a_1|\Phi_0\rangle_1|\Psi_0\rangle_2, \qquad (23)$$

is generated provided that vacuum (no-click) is registered in the auxiliary mode. As, according to Table 1, the parities of the CV components of the state $|\Omega_0\rangle$ are different ($|\Psi_0\rangle$ is even and $|\Phi_0\rangle$ is odd), the conditional state is also described in four-dimensional Hilbert space. The state can no longer be called hybrid since its components are exclusively CV ones living in infinite Hilbert space. We are going to call the states like in equation (23) CV entangled states. Generation of such type of the CV entangled state can greatly expand the capabilities of the CV quantum information processing on analogy with entangled coherent state consisting of coherent states with opposite in sign amplitudes [19]. If a balanced delocalized photon has been pre-selected then the output state (23) becomes maximally entangled.

Generation of the conditional states can be continued further depending on the measurement outcomes in the auxiliary third mode

$$|\Omega_{2m}\rangle_{12} = N'_{2m}\left(a_0|\Psi_0\rangle_1|\Phi_{2m}\rangle_2 - a_1\frac{B_0 r K_{2m}}{2m L_{2m}}|\Phi_0\rangle_1|\Psi_{2m}\rangle_2\right), \tag{24}$$

provided that $p = 2m$ Fock state is detected and

$$|\Omega_{2m+1}\rangle_{12} = N'_{2m+1}\left(a_0|\Psi_0\rangle_1|\Phi_{2m+1}\rangle_2 - a_1\frac{B_0 r K_{2m+1}}{(2m+1)L_{2m+1}}|\Phi_0\rangle_1|\Psi_{2m+1}\rangle_2\right), \tag{25}$$

if $p = 2m + 1$ number state is registered. Here, $N'_{2m}$ and $N'_{2m+1}$ are the corresponding normalization factors. The degree of the entanglement of the states can also be estimated using the expression in Eq. (22), where now one should use modified additional parameters $B'_{2m} = (B_0 r K_{2m})/(2m L_{2m})$ and $B'_{2m+1} = (B_0 r K_{2m+1})/((2m+1)L_{2m})$. Obviously, one can manipulate the initial parameters $(r, t, |a_1|)$ in order to adapt them so that negativity of the states in Eqs. (24)-(25) takes on the maximum possible value. An entangled state with varying degrees of entanglement is generated for any measurement outcome in the auxiliary mode, therefore, the generation can also be called receipt of the CV entangled states on demand.

In conclusion, we have developed a new way of generating entangled states, both hybrid and pure CV. Any generated state has a certain degree of entanglement. Knowing the initial values of the parameters $(r, t, |a_1|)$ and the measurement outcome, one can always estimate the negativity of the output state. The following physical explanation of the generation can be given. If even number of photons comes from even SMSV in Eq. (1), then heralded state will also contain even number of photons provided that even number of photons is detected at auxiliary mode. If instead, even number of photons of the even SMSV is mixed with single photon of the delocalized photon, the resulting state will contain odd number of photons in the case of registration of even number of photons at auxiliary mode. Due to indistinguishability between the two events, conditional either hybrid in Eqs. (7), (12) or CV in Eq. (24) state is obtained. The CV components of the heralded entangled state acquire the parity as noted at second column of the Table 1. The same explanation holds in the case of detecting odd number of photons in the auxiliary mode. What is interesting is that this mechanism allows us to introduce new CV states and thereby expand the boundaries of quantum state engineering. The number of the optical elements to implement the entanglement is minimal. The resources used to implement the entanglement are available, especially regarding routine used in experiments single mode squeezed vacuum. A single photon is generally obtained conditioned and requires a certain procedure, but nevertheless, a single photon is also achievable by modern technologies. Thus, the experiment is implementable into practice. This mechanism can be expanded to create more complex states (quantum networks) from a large number of modes that would contain both DV and CV states. Moreover, the quantum network can also be implemented from SMSV and single photons in deterministic manner.

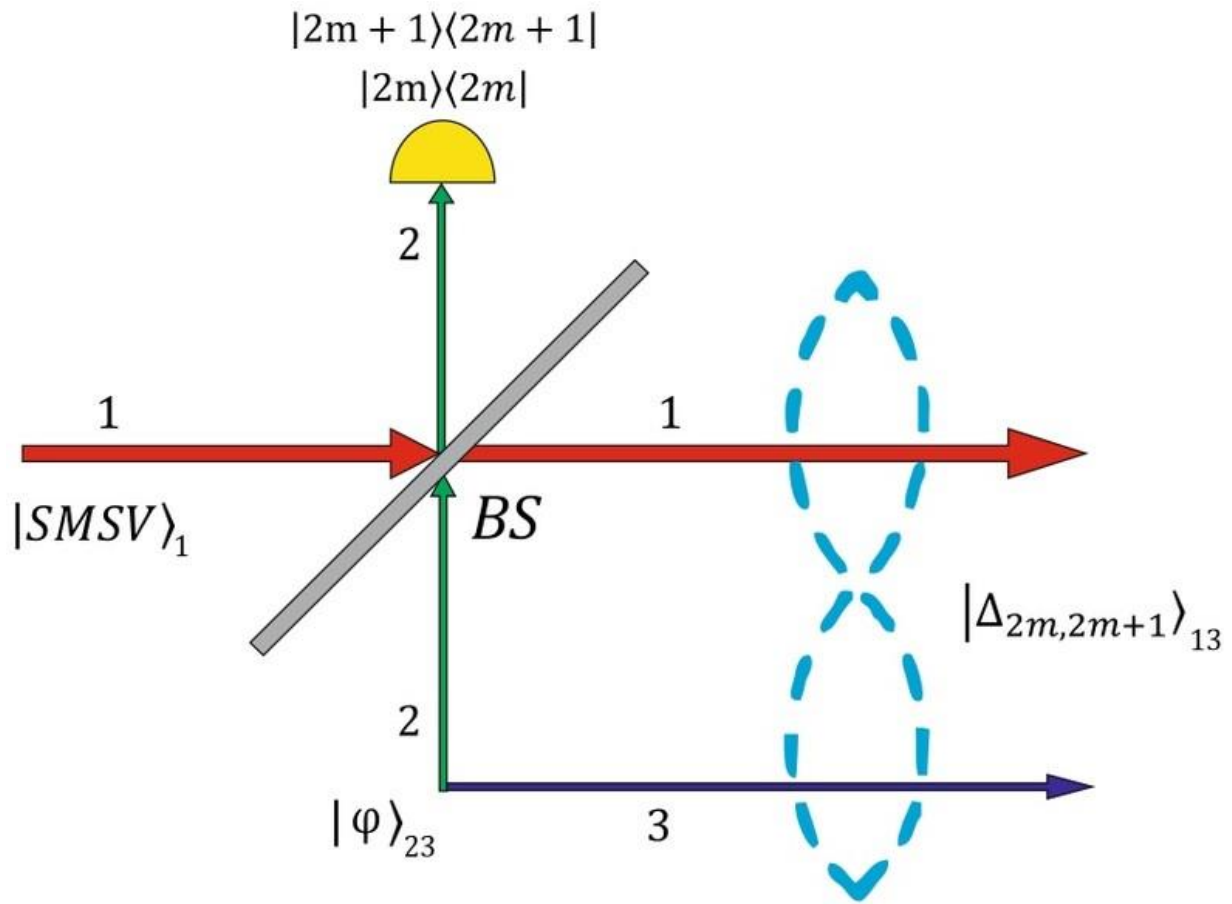

**Fig. 1.** Optical scheme for the conditional generation of the hybrid entangled states either $|\Delta_{2m}\rangle_{13}$ or $|\Delta_{2m+1}\rangle_{13}$. SMSV interacts with second mode of the delocalized photon with any amplitudes followed by registration of the measurement outcome in the same mode. The parameters of the beam splitter are arbitrary. The squeezing parameter $r$ of the SMSV can be chosen arbitrary. There are values of the parameters at which the output state has maximum entanglement. Also, the scheme can be adapted for conditional generation of CV entangled states (23)-(25).

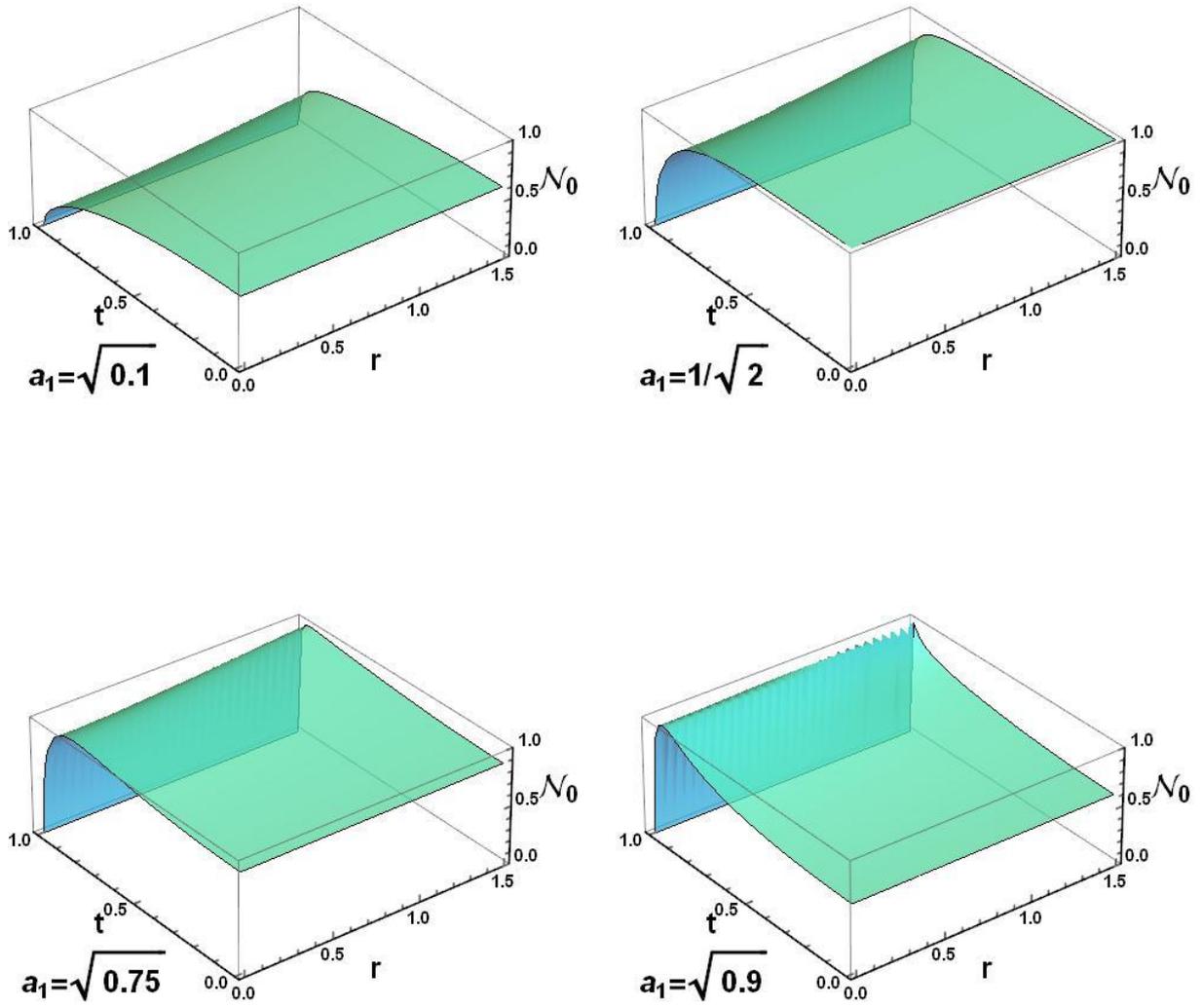

**Fig. 2.** Three-dimensional plots of the negativity $\mathcal{N}_0$ of the conditional hybrid entangled state $|\Delta_0\rangle$ in dependency on squeezing parameter $r$ of the SMSV and transmittance of the beam splitter $t$ for different values of $|a_1|$ characterizing the nonlocality of the input single photon.

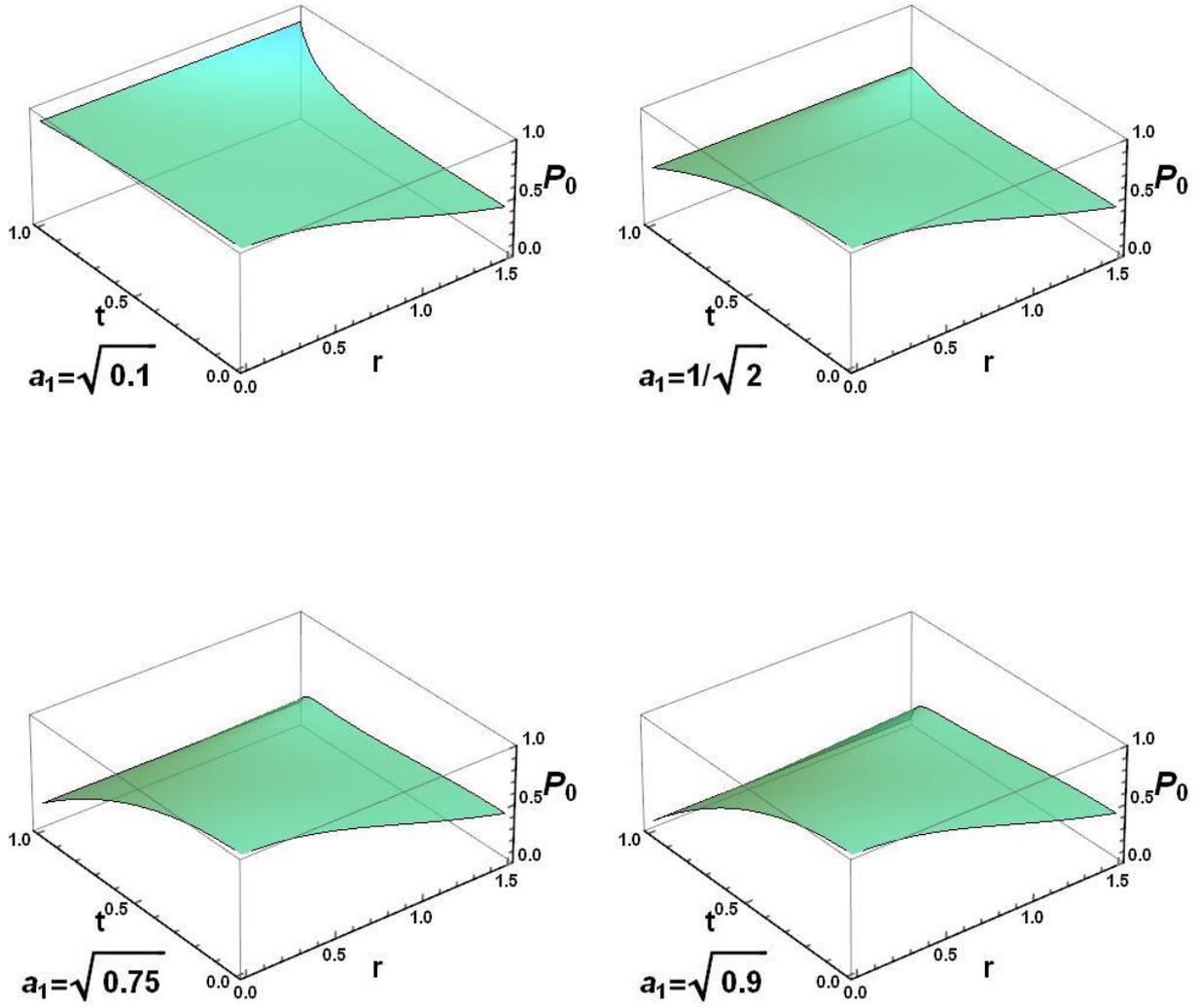

**Fig. 3.** Three-dimensional plots of the success probability $P_0$ of the conditional hybrid entangled state $|\Delta_0\rangle$ in dependency on squeezing parameter $r$ of the SMSV and transmittance of the beam splitter $t$ for different values of $|a_1|$ characterizing the nonlocality of the input single photon.